# A Decompilation Approach to Partitioning Software for Microprocessor/FPGA Platforms

Greg Stitt and Frank Vahid*
Department of Computer Science and Engineering
University of California, Riverside
{gstitt, vahid}@cs.ucr.edu, http://www.cs.ucr.edu/{~gstitt, ~vahid}
*Also with the Center for Embedded Computer Systems, UC Irvine

## Abstract

*In this paper, we present a software compilation approach for microprocessor/FPGA platforms that partitions a software binary onto custom hardware implemented in the FPGA. Our approach imposes less restrictions on software tool flow than previous compiler approaches, allowing software designers to use any software language and compiler. Our approach uses a back-end partitioning tool that utilizes decompilation techniques to recover important high-level information, resulting in performance comparable to high-level compiler-based approaches.*

## 1. Introduction

Several commercial platforms have begun to integrate microprocessors and FPGA onto a single-chip. Designers have typically used the FPGA in these platforms for implementing peripherals. More recently, designers have used the FPGA to implement custom hardware that speeds up the execution of software running on the microprocessor.

For several platforms, platform vendors have developed compilers to enable software developers to automatically partition high-level software source code onto the FPGA in order to speedup software execution. These compilers provide good technical solutions for partitioning software but impose several restrictions that software developers may find undesirable. One restriction is that these compilers generally support only one particular language, generally C/C++. Software developers also generally have well-established software development tools and would resist a change to a different compiler.

Thus, a partitioning tool would likely be far more practical if placed after the compiler in the tool flow, operating during the software linking stage, or by parsing the final software binary. The partitioning/synthesis tool would thus be independent of the compiler tool – any source language and/or compilers could be utilized. The partitioning/synthesis tool could be provided by the platform vendor, thus incorporating specific knowledge of the platform architecture, which may involve extensive details involving communication, memory, interrupt, arbitration, etc., necessary to perform good partitioning and synthesis. Such an approach will not replace the use of specialized languages and compilers used by advanced designers, but will instead extend the advantages of microprocessor/FPGA platforms to a much wider range of software developers, for whom the possibly lower-quality results compared to a compiler approach are still far better than software-only implementations.

In this paper, we show that the key to performing good partitioning and synthesis after compilation is to be able to recover the necessary high-level information. Such recovery is known as decompilation. We have therefore developed an extensive set of decompilation methods specifically intended for partitioning and synthesis. We point out that our methods are also applicable for synthesizing an entire software application, not just kernels, to a custom circuit.

## 2. Decompilation

Decompilation was originally developed for purposes of translating software binaries from one instruction set architecture to another and for recovering high-level code from legacy assembly code. We use decompilation for a different purpose, namely for converting a software binary into a representation suitable for synthesis. That different purpose meant we had to select among existing decompilation methods and adapt them to our needs, and also that we had to develop new decompilation methods specifically for our purpose.

Our decompilation process uses existing decompilation techniques [1] to convert the software binary into a control/data flow graph (CDFG) that is annotated with high-level information. Initially, *binary parsing* converts the software binary into an instruction set independent representation. Next, *CDFG creation* builds a control/data flow graph (CDFG) for the application. *Control structure recovery* analyzes the CDFG and determines high-level control structures, such as loops and if statements.

After recovering a CDFG of the application, we apply several optimizations to eliminate overhead introduced by the instruction set. One such overhead is the use of arithmetic instructions with a immediate value of zero in order to move a value between two registers. Although a move instruction is more appropriate, binary-level synthesis cannot assume a compiler will use instructions appropriate for synthesis. If the arithmetic operator is synthesized, then large amounts of area will be wasted. We remove this overhead using constant propogation. We also perform operator size reduction, strength reduction, and stack operation removal.

In addition to removing instruction set overhead, we must also undo software compiler optimizations in order to make the recovered CDFG more appropriate for synthesis. Strength reduction of multiplication operations is a compiler optimization that can reduce the quality of binary-level synthesis. Although strength reduction is generally beneficial, the additional adders and shift resources required to perform multiplications may exhaust these resources, leading to increased latency, even if hardware multipliers are available. To achieve the fastest hardware, the synthesis tool must decide whether strength reduction is beneficial. To give the synthesis tool this added flexibility, we perform strength promotion to convert series of shift/add operations back into the original multiplication form.

Loop unrolling can also result in inefficient binary-level synthesis. Loop unrolling can obscure high-level information such as memory access patterns and resource requirements, which are needed for effective synthesis. Loop unrolling can also greatly increase the size of a software binary, which can increase synthesis execution times and memory requirements, making dynamic synthesis approaches infeasible. We use loop rerolling to identify unrolled loops and then roll the loops back into a representation similar to their original representation in high-level code.

## 3. Partitioning and synthesis

Although we considered using standard hardware/software partitioning approaches [2][3], we use a simpler technique based on the well-known 90-10 rule in order to reduce the time required for partitioning. Achieving a small partitioning execution time is important because we intend to integrate our approach with existing dynamic partitioning and dynamic synthesis approaches [4]. Our partitioning algorithm proceeds in three steps. In the first step, we use profiling results to identify the most frequent few loops, which generally correspond to 90 percent of execution while consisting of only a few dozen lines of code. We then include these loops in the hardware partition. In the second step, we use alias information to find regions of code that access the same memory locations as the loops in the hardware partition. If space allows, we include these regions in the hardware partition so that the required memory locations can be moved to memory within the FPGA, increasing parallelism. In the third step, we continue to add regions to the hardware partition based on profiling results and hardware suitability until the area constraint is violated. This final step allows an entire application to be synthesized if space allows.

Our approach utilizes a behavioral synthesis tool that we implemented ourselves. The input to the synthesis tool is the decompiled CDFG for the regions selected for hardware implementation. The output of the tool is register transfer-level VHDL. We use Xilinx ISE to synthesize the VHDL to a netlist.

## 4. Results of decompilation-based partitioning

We applied our decompilation-based partitioning approach to twenty examples from EEMBC, PowerStone, MediaBench, and our own benchmark suite. All examples were compiled using gcc with –O1 optimizations.

Instead of using a commercial platform, we utilized a hypothetical platform consisting of a MIPS microprocessor and Xilinx Virtex II FPGA. Using a hypothetical platform allows us to more easily evaluate different types of platforms with different clock speeds and FPGA sizes.

The decompilation-based approach showed consistently good application speedups and energy savings, averaging 5.4 and 69%, compared to a MIPS processor running at 200 MHz. The average kernel speedup was 44.8. Compared to a 400 MHz MIPS, the application speedups were 3.8 and the energy savings were 49%. For slower platforms with a 40 MHz microprocessor, the application speedup was 12.6 and the energy savings were 84%. The average area required was an equivalent of 26,261 logic gates. For these examples, our approach recovered almost *all* the relevant high-level constructs successfully. The only unsuccessful situations occurred during CDFG recovery, which failed for two EEMBC examples because of indirect jumps.

In addition to the experiments based on software binaries generated with –O1 optimizations, we performed the same experiments on binaries generated using four different optimizations levels for four of the previous examples. As expected, software execution times improved as the level of compiler optimizations increased. In most cases, the execution times of the synthesized examples also improved with more compiler optimizations. This phenomenon implies that software compiler optimizations generally do not negatively impact binary-level synthesis, and in many cases binary-level synthesis actually improves when more compiler optimizations are applied. Speedup was significant for all levels of compiler optimizations, although the speedup did not always increase with more compiler optimizations. Speedup did not always increase because as more compiler optimizations are applied, the software became significantly faster, which increased the difficulty of achieving large speedups. Note that execution time is the true measure of the effectiveness of binary-level synthesis. We report speedups to simply show that significant improvements are achieved for each individual level of compiler optimization. The energy savings were also very similar across different levels of compiler optimizations.

## 5. Acknowledgements

This research was supported in part by the National Science Foundation (CCR-0203829) and by the Semiconductor Research Corporation (2003-HJ-1046G).